# Multi mode nano scale Raman echo quantum memory

S.A.Moiseev[1,2*] and E.S.Moiseev[1]

[1)] *Kazan Physical-Technical Institute of the Russian Academy of Sciences*
*10/7 Sibirsky Trakt, Kazan, 420029, Russia;*
[2)] *Institute for Informatics of Tatarstan Academy of Sciences,*
*20 Mushtary, Kazan, 420012, Russia*
\* E-mail: samoi@yandex.ru.

**Abstract**

Low loss magnetic surface plasmon polariton (SPP) modes characterized by enhanced electrical field component and subwavelength confinement on the dielectric and negative-index metamaterial interface are presented. We demonstrate a possibility of storage and perfect retrieval of the low loss magnetic SPP fields by using a photon echo quantum memory on Raman atomic transition. We describe specific properties of the proposed technique which opens a possibility for efficient nano scale multi-mode quantum memory.

**Introduction**

Construction of quantum computer and realization of quantum communication along large distances request optical quantum memory (QM) as one of the key elements of the basic quantum information processes [1-3]. Besides, present technologies require the quantum devices integrated to the nano scale schemes that determines a long-term challenge for engineering of the quantum computer hardware [4-7]. A considerable success has been achieved recently in optical QM [8-11] however its realization in the nano scale circuits remains almost intact problem of theoretical and experimental investigations.

Recognized methods that allow overcoming physical problems of the light field confinement in subwavelength dimensions include exploiting photonic crystals [12] or surface polaritons [13]. Recently we proposed a nano scale coherent control of photons via its transfer to the surface plasmon polariton (SPP) fields on dielectric and negative-index metametrial (NIMM) interface (D/NIMM-interface) and resonant interaction with coherent atomic systems localized on the interface [14]. NIMM is artificially fabricated medium characterized by homogeneous permittivity $\varepsilon < 0$ and permeability $\mu < 0$. A number of interesting optical properties occur with a light in the NIMMs that are principally impossible in usual media [15]. Recently we observed that the D/NIMM interface provides a convenient possibility to realize a transverse subwavelength confinement with losses suppression of the SPP fields [14] that could be interesting for enhancement of the nonlinear photon-photon interactions in the nano scale scheme [16]. In this work we describe a detailed physical picture of subwavelength confinement accompanied by the low losses of the transverse magnetic SPP modes for D/NIMM interface with $\mu_2 \approx 0$ and then we demonstrate a multi mode nano scale optical QM of the SPP field on the D/NIMM interface. Here we use Raman echo QM technique proposed recently in [17, 18]. Finally we discuss the obtained results with promising possibilities of the nano scale QM.

**Subwavelength low loss transverse magnetic surface plasmon polaritons**

D/NIMM -interface contains two media: a dielectric with positive $\varepsilon_1 > 0, \mu_1 > 0$ in upper plane z>0 and the NIMM with $\varepsilon_2 = \varepsilon_r + i\varepsilon_i$ ($\varepsilon_r < 0$, $\varepsilon_i > 0$) and $\mu_2 = \mu_r + i\mu_i$ ($\mu_r \leq, \mu_i > 0$) for down plane z<0. Resonant atoms are situated in within thin layer $0 < z < z_o$ with density $n_o$ (for example it could be NV centers in the diamond, or $Pr^{3+}$ ions doped in YSiO crystal) that can be fabricated with recently proposed technology [19]. Below we use resonant atoms for QM of the SPP fields propagating along x-direction on the D/NIMM interface. As we found numerically in [14], the SPP fields acquire reduced losses in wide frequency range around fixed frequency $\omega_o$ where the magnetic permeability $\mu_r$ is close to zero. The frequency $\omega_o$ depends on the real and imaginary parts of the dielectric and NIMM permittivities and permeabilities. By taking into account weaker magnetic interactions we ignore magnetic losses of SPP field in the NIMM by assuming $\mu_i = 0$.

Let us evaluate the SPP field properties in the frequency range where $\mu_2 \approx 0$ so that a magnetic induction of the SPP field is close to zero ($\vec{B}_2 \approx 0$) in the NIMM. In accordance with the boundary conditions, the surface field components of the transverse magnetic SPP modes satisfy to the relations: $H_y^{(1)}(z=0) = H_y^{(2)}(z=0) = \omega\varepsilon_1 E^{(1)}/(ck_1^p)$, $\vec{E}_x^{(1)} = \vec{E}_x^{(2)}$ and the normal components $\vec{E}_z^{(2)} = (\varepsilon_1/\varepsilon_2)\vec{E}_z^{(1)}$, $\vec{H}_z^{(1)} = \vec{H}_z^{(2)} = 0$ (where $k_j^p = \sqrt{(K_\parallel^p)^2 - \omega^2\varepsilon_j\mu_j/c^2}$ for j=1,2). By taking into account $\mu_2 \to 0$ we find the complex wavevector of the SPP modes

$$K_\parallel^p = k_\parallel^p + i\kappa^p\big|_{\mu_2 \to 0} = \frac{2\pi}{\lambda_o\sqrt{1-(\varepsilon_1/\varepsilon_2)^2}}, \tag{1}$$

where $\lambda_o = c/(2\pi\omega\sqrt{\varepsilon_1\mu_1})$ is a wavelength of free light modes in 3D dielectric, $\kappa$ is an absorption coefficient, wavelength of SPP mode along x-direction is $\lambda_\parallel = 2\pi/k_\parallel$.

The transverse confinement of the probe (*p*-) SPP modes is determined by exponential factors: $\vec{E}_p^{(1)}(z) = [\vec{e}_x + i\vec{e}_z K_\parallel^p/k_1^p]\exp\{-k_1^p z\}E_o(k_\parallel)$ in z>0 where $E_o(k_\parallel) = \sqrt{\hbar\omega(k_\parallel)/2\pi\varepsilon_o L_y L_z}$ is an electric field of SPP quanta and $\vec{E}_p^{(2)}(z) = [\vec{e}_x - i\vec{e}_z K_\parallel^p/k_2^p]\exp\{k_2^p z\}E_o(k_\parallel)$ in z<0, where $k_1^p/k_2^p = -\varepsilon_1/\varepsilon_2$ for transverse magnetic SPP modes, $L_y$ and $L_z$ are the quantization lengths of SPP modes, $\varepsilon_o$ is a vacuum permittivity [16].

Let us assume the following inequality for imaginary part of the electric permittivity

$$\varepsilon_i/|\varepsilon_r| \ll |\varepsilon_r + \varepsilon_1|/|\varepsilon_r| \ll 1, \qquad (2)$$

which means that we stay at some spectral distance from the point $|\varepsilon_r(\omega)| \cong \varepsilon_1$. Here, the losses will cause a negligible influence on the SPP mode structure. By using (2) in (1) we find

$$k_\parallel^p + i\kappa^p\big|_{\mu_2 \to 0} \cong \frac{2\pi}{\lambda_o}\sqrt{\frac{\varepsilon_1}{2\varepsilon_i}}\sqrt{\frac{u}{1-iu}}, \qquad (3)$$

where $u = 2\varepsilon_i|\varepsilon_r|/(|\varepsilon_r|^2 - \varepsilon_i^2 - \varepsilon_1^2)$. From (3) we find the wavelength $\lambda_\parallel$ and propagation length $l_x^p = 1/\kappa^p$ of SPP pulse along the D/NIMM interface (see Fig. 1). The length $l_x^p$ and $\lambda_\parallel$ become close to each other for $u \gg 1$ where highest confinement coexists with increasing propagation losses $\kappa^p$. We find a possibility of high confinement accompanied by weak losses for the following relation between $\varepsilon_i$ and $\varepsilon_1$: $\varepsilon_i/|\varepsilon_r| \approx 10^{-4}$ and $|\varepsilon_r + \varepsilon_1|/|\varepsilon_r| \approx 10^{-2}$. Here, $k_\parallel^p + i\kappa^p \cong 2\pi\sqrt{50}\{1 + i/200\}/\lambda_o$ which leads to large transverse subwavelength confinement $\xi_1^p = 1/k_1^p = \lambda_o/(14\pi) \approx \lambda_o/40$ for $z > 0$ ($\xi_1^p/\lambda_\parallel = 1/2\pi \approx 0.16$) which is comparable with the confinement $\xi_2^p = 1/|k_2^p| = \lambda_o/(2\sqrt{50}\pi) \cong \xi_1^p = \xi$ for $z < 0$ (see Fig.2,3). It is most important, that the SPP modes are characterized by large propagation length in comparison with the transverse confinement $l_x/\xi \geq 150$ ($l_x = 1/\kappa = 4.5\lambda_o$) so we get low loss subwavelength SPP modes for a distance $L_x \ll l_x$.

Before analysis of QM on the subwavelength SPP modes, we compare its magnetic and electric components with the components of free propagating light modes in the NIMM volume with $\varepsilon_r < 0$, $\mu_2 \approx 0$. The SPP fields are satisfied to the ratio $H_y^{(2)}/E_o \cong H_y^{(1)}/E_o = (2\pi\xi_1^p/\lambda_o)\sqrt{\varepsilon_1/\mu_1}$ which shows a suppression of the magnetic field component by factor $(2\pi\xi_1^p/\lambda_o) = 1/7 \approx 0.14$ with respect to the ratio of the free light modes in 3D dielectric. In contraposition to SPP, the free propagating fields acquire enhanced magnetic component in 3D NIMMs in accordance with general relation $E/H = \sqrt{\mu_2/\varepsilon_2}\big|_{\mu_2 \to 0} \to 0$. Thus the transverse magnetic SPP modes have a subwavelength confinement and enhanced electric field component near the D/NIMM interface that is very important for the interaction of the SPP fields with atomic systems characterized by the electrical dipole transitions. For complete spectral characterization of the transverse magnetic SPP-field we take into account usual spectral behavior of $\varepsilon_2, \mu_2$ [15]

$$\varepsilon_2 = \varepsilon_o - \frac{\omega_e^2}{\omega(\omega + i\gamma_e)}, \quad \mu_2 = \mu_o - \omega_\mu^2/\omega^2, \tag{4}$$

where $\omega_e = 1.37 \cdot 10^{16}$, $\varepsilon_o = \mu_o = 2$, $\gamma_e$ is a decay constant. By putting $\mu_2 = 0$ in (4) we find the transverse quantization length $L_z$ of the SPP field [16] as a nonlinear function of spatial confinements $\xi_1^p$ and $\xi_2^p$:

$$L_z = \{2\varepsilon_1 + \varepsilon_1(2\pi\xi_1^p/\lambda_o)^2\}\xi_1^p + \{2(2\varepsilon_o + \varepsilon_1) + 2\varepsilon_1(2\pi\xi_p^p/\lambda_o)^2 \frac{\mu_o}{\mu_1}\}\xi_p^p. \tag{5}$$

$L_z$ will be a linear function of the confinement increased at $|\varepsilon_r(\omega)| \to \varepsilon_1$ in accordance with (1)

$$L_z = 4(\varepsilon_o + \varepsilon_1)\xi = 4\mu_o(\omega_e^2/\omega_\mu^2)\xi = 0.55\lambda_o. \tag{6}$$

where $\xi_2^p \cong \xi_1^p = \xi \cong \lambda_o/40$, $\omega_e = 1.67\omega_\mu$, $\mu_1 = 1$. Below we describe a nano scale QM for the magnetic SPP modes. At first we motivate a choice of the Raman echo QM (REQM) technique and then we describe its main properties.

**Raman echo quantum memory of surface plasmon polariton fields**

REQM [17,18] is a generalization to the Raman transitions of the photon echo QM protocol proposed originally for gases in [20,21] and then was adopted to the solid state systems in [22-25] and extended in [26] to the mechanism of passive rephasing of the atomic coherence. Number of stored modes in the photon echo QM techniques is limited only by ratio of the inhomogeneous to the homogeneous broadening of the resonant atomic transition $\sim \Delta_{in}/\gamma$ that usually outnumbers about $10^2$-$10^4$. It was recently found [17] that REQM eliminates some critical limitations in realization of reversible light-atoms dynamics inhering to the photon echo QM protocol. In particular the REQM can work within wider spectral range and can be exploited for efficient wavelength conversion of the quantum light fields. Another important advantage of the REQM is a direct transfer of the quantum information from the probe light field on the long-lived atomic transitions without using the control laser π-pulses. Here we show how such QM works in nano scale circuits where the SPP pulses are characterized by highly inhomogeneous spatial field profile (see Fig.1) so the control π-pulses are impossible for realization on the D/NIMM interface.

We assume that initially all atoms are prepared on level 1 and a population of excited levels 2 and 3 will be negligible due to the interaction with weak probe SPP pulse (characterized by index "$p$"). Then we launch the weak probe light pulse on the D/NIMM interface where the pulse transfers in SPP field $A_p(r,t)$ propagating along $\vec{K}_p \parallel \vec{x}$-direction at the condition of negligible low losses at a distance $L_x \ll l_x$ (where $\vec{K}_p = k_\parallel(\omega_1)\vec{e}_x$). We also assume that the probe quantum field $A_p(r,t)$ is excited in the presence of additional intensive control classical field $\Omega_{cp}(\vec{r},t)$ propagating along $\vec{K}_{cp}$ with small angle deviation from x-axis as depicted in Fig.4. $\hat{A}_p(t,x)$ and $\Omega_{cp}(\vec{r},t)$ fields are characterized by the transverse confinement factors $\xi_1^p = 1/k_1^p$ and $\xi_1^{cp} = 1/k_1^{cp}$ in the plane z>0 (where $\hat{A}_p(t,x) = \exp\{i\omega_p t - iK_p x\}\int dk_\parallel (\vec{e}_1 \vec{E}_{0a}^p(k_\parallel))e^{ik_\parallel x}\hat{a}(k_\parallel,t)$ [14], $\hat{a}(k_\parallel,t)$ and $\hat{a}^+(k_\parallel',t)$ are the bosonic operators $[\hat{a}(k_\parallel,t),\hat{a}^+(k_\parallel',t)] = \delta(k_\parallel - k_\parallel')$, $\vec{E}_{0a}^p(k_\parallel) = [\vec{e}_x + i\vec{e}_z K_\parallel^p/k_1^p]E_o(k_\parallel)$). The two fields $\hat{A}_p(t,x)$ and $\Omega_{cp}(\vec{r},t)$ induce a Raman transition between atomic level 1→2 in the layer $0<z<z_o$ of the D/NIMM interface as depicted in Fig.5. By generalizing the approach [14,16,17] to the nano scale scheme, we derive the system of "P-" equations for the SPP field and three-level atomic coherence in the Heisenberg picture (in moving coordinate system $X = x$, $\tau = t - x/\upsilon_o$).

$$\frac{\partial}{\partial X}\hat{A}_p(\tau,X) = i\sum_j \exp\{-k_1^p(\omega_p)z_j\}M_p(z_j, X - x_j)\hat{S}_{p,13}^j(\tau), \tag{7}$$

$$\frac{\partial}{\partial \tau} \hat{S}^{j}_{p,13} = -i(\Delta^{j}_{31} + \Delta_{p})\hat{S}^{j}_{p,31} + i\exp(-k^{p}_{1}z_{j})(\vec{e}_{p}\vec{d}_{13}/\hbar)\hat{A}_{p}(\tau, X_{j}) + i\exp(-k^{cp}_{1}z_{j})\Omega_{cp}(\tau)\hat{S}^{j}_{p,12}, \quad (8)$$

$$\frac{\partial}{\partial \tau} \hat{S}^{j}_{p,12} = -i(\Delta^{j}_{21} + \Delta^{R}_{p})\hat{S}^{j}_{p,12} + i\exp(-k^{cp}_{1}z_{j})(\Omega_{cp}(\tau))^{*}\hat{S}^{j}_{p,13}, \quad (9)$$

where $\Delta_{p} = \omega_{31} - \omega_{p}$ is detuning from the optical resonance, $\Delta^{j}_{21}$ and $\Delta^{j}_{31}$ are the frequency detunings of the atomic transitions 1→2 and 1→3, $\Delta^{R}_{p} = \omega_{21} - \omega_{cp} + \omega_{p}$ is a detuning from Raman resonance, $\omega_{p}$ and $\omega_{cp}$ are the carrier frequencies of the probe and first control fields, $\vec{e}_{p}$ is a polarization vector, $\hat{S}^{j}_{p,12} = \hat{P}^{j}_{12} \exp\{-i(\omega_{cp} - \omega_{p})t + i(\vec{K}_{cp}\vec{r}_{j} - K_{p}x_{j})\}$, $\hat{S}^{j}_{p,13} = \hat{P}^{j}_{13} \exp\{i\omega_{p}t - iK_{p}x_{j}\}$, $\hat{P}^{j}_{m'm} = |m'\rangle_{j}\langle m|$ are the atomic operators, $\vec{d}_{m'm}$ and $\omega^{j}_{m'm}$ are the dipole moment and frequency of the transition m'→m for j-th atom. It is worth noting that the imaginary part of $M_{p}(z_{j}, X - x_{j})$ equals zero for spatially homogeneous media. For the optical wavelengths we can use usual ansaz

$$M_{p}(z_{j}, X - x_{j}) \cong 2\pi \frac{(\vec{e}_{p}\vec{E}^{p}_{0a}(k_{\parallel}))(\vec{d}_{13}\vec{E}^{p}_{0a}(k_{\parallel}))^{*}}{v_{g}\hbar}\delta(X - x_{j}). \quad (10)$$

where $v_{g} = \partial\omega/\partial k_{\parallel}$ is a group velocity of the SPP pulse without resonant atoms.

By assuming a large enough spectral detuning of the probe field from the atomic frequencies of optical transition $|\Delta^{j}_{31} + \Delta_{p}| \gg \delta\omega$ ($\delta\omega$ is a spectral width of the probe field) we exclude the excited atomic coherence $\hat{S}^{j}_{p,31}$ in (6)-(8) and obtain two coupled equations for $\hat{A}_{p}(\tau, X)$ and long lived coherence $\hat{S}^{j}_{p,12}$. Evolution of the input field is expressed via its Fourier image $\hat{A}_{p}(\tau, X) = \int d\nu \exp\{-i\nu\tau\}\hat{A}_{p,\nu}(X)$ where

$$\hat{A}_{p,\nu}(X) = \exp\{[i\varsigma_{p} - \tfrac{1}{2}\alpha_{p}(\nu)]X\}\hat{A}_{p,\nu}(0), \quad (11)$$

$$\varsigma_{p} = \chi_{p}\left\langle\frac{\exp\{-2k^{p}_{1}z_{j}\}}{(\Delta^{j}_{31} + \Delta_{p} - i\gamma_{31})}\right\rangle_{j}, \quad (12)$$

$$\alpha_{p}(\nu) = 2\chi_{p}\left\langle\frac{1}{[i(\delta^{j}_{p} - \nu) + \gamma_{21}]}\frac{\exp\{-2(k^{p}_{1} + k^{cp}_{1})z_{j}\}}{(\Delta^{j}_{31} + \Delta_{p} - i\gamma_{31})^{2}/|\Omega_{cp}|^{2}}\right\rangle_{j}, \quad (13)$$

$$\langle F(\Delta^{j}_{21}, \Delta^{j}_{31}, z_{j})\rangle_{j} = \int_{0}^{z_{o}} dz_{j} \int_{-\infty}^{\infty} d\Delta^{j}_{21} \int_{-\infty}^{\infty} d\Delta^{j}_{31} G(\Delta^{j}_{21}, \Delta^{j}_{31})F(\Delta^{j}_{21}, \Delta^{j}_{31}, z_{j}), \quad (14)$$

where we introduced the decay constants $\gamma_{21}$ and $\gamma_{31}$ which can be ignored for interaction with a short SPP probe pulse, $G(\Delta_{21}^j, \Delta_{31}^j)$ is a function of inhomogeneous broadening of the atomic transitions 1→2 and 1→3, $\chi_p = 2\pi n_o L_y |(\vec{d}_{13}\vec{E}_{0a}^p)|^2/\upsilon_o\hbar^2$, $<F_j..>_j$ describes a total atomic response in the cross-section ($L_y \times z_o$) of the layer 0<z<z_o determined by function F

$$\delta_p^j = \Delta_{21}^j + \Delta_p^R - \frac{\exp\{-2k_1^{cp}z_j\}|\Omega_{cp}|^2}{(\Delta_{31}^j + \Delta_p - i\gamma_{31})}. \tag{15}$$

The atomic detuning $\delta_p^j$ of Raman transition depends on $z_j$ that is caused by exponential dependence of the Rabi frequency on the spatial distance from the D/NIMM interface.

Let us proceed step by step to the main requirements of the perfect QM realization. In accordance with (10), the perfect storage of a single or multi-mode probe field occurs at the complete absorption of SPP field if the condition $\alpha(\nu)L_x \gg 1$ is satisfied for all its spectral components $\nu \subseteq \delta\omega$. By assuming the perfect storage we examine a retrieval of the echo signal irradiated to the backward "-x" direction after the additional control laser pulse will be applied. The control SPP pulse propagates in $K_{ec}$-direction with carrier frequency $\omega_{ec}$ (see Figs. 4,5) and is characterized by the Rabi frequency $\Omega_{ec}$. In this case we get new "E-" system of equations where the echo signal irradiation is given by

$$-\frac{\partial}{\partial X}\hat{A}_e(\tau_e, X) = i\sum_j \exp\{-k_1^e(\omega_e)z_j\}M_e(z_j, X-x_j)\hat{S}_{e,13}^j(\tau_e). \tag{16}$$

The echo signal propagates along wave vector $\vec{K}_e = -K_e\vec{e}_x$ with new carrier frequency $\omega_e$ (see Figs.4,5), where $X = x, \tau_e = t + x/\upsilon_o$, and $\xi_1^e = 1/k_1^e$, $\xi_1^{ec} = 1/k_1^{ec}$ characterize a transverse confinement of the echo signal and second control fields.

We find the material equations for $\hat{S}_{e,12}^j = \hat{P}_{12}^j \exp\{-i(\omega_{ec}-\omega_e)t + i(\vec{K}_{ec}\vec{r}_j + K_e x_j)\}$ and $\hat{S}_{e,13}^j = \hat{P}_{13}^j \exp\{i\omega_e t + iK_e x_j\}$ of "E"-equations by replacing the indexes "p"→"e" and $\tau \to \tau_e$ in all variables and parameters of "P"-equations. By comparing the evolution of values ($\hat{A}_p(\tau, X)$, $\hat{S}_{p,12}^j$, $\hat{S}_{p,13}^j$) in "P"-system with the evolution of ($\hat{A}_e(\tau_e, X)$, $\hat{S}_{e,12}^j$, $\hat{S}_{e,13}^j$) accordingly in "E"-system, we find that the two systems of equations coincide with each other if we will reverse a time arrow of the echo signal evolution at t' as $\tau_e \to -\tau_e$ and perform the following transformations in the "E"-equations.

1) Transfer to new field light field amplitude $\hat{A}_e(\tau_e, X) \to -\hat{A}_e(\tau_e, X)$.

2) Transverse confinement for the quantum and control fields are the same: $\xi_1^p(\omega_p) = \xi_1^e(\omega_e)$ and $\xi_1^{cp}(\omega_{cp}) = \xi_1^{ce}(\omega_{ce})$. Moreover also we can take $\xi_1^e(\omega_e) = \xi_1^{ce}(\omega_{ce}) = \xi$ for $\omega_{21}/\omega_{31} \ll 1$ and $|(\omega_p - \omega_{cp})/\omega_p| \approx |(\omega_e - \omega_{ep})/\omega_e| \ll 1$.

3) We are able to inverse the atomic detunings of the inhomogeneous broadenings for the both atomic transitions 1-3, 1-2: $\Delta_{31}^j(t > t') = -\Delta_{31}^j(t < t')$, $\Delta_{21}^j(t > t') = -\Delta_{21}^j(t < t')$ (that is an usual CRIB procedure) and to change a carrier frequency of second control field so that $\Delta_e = -\Delta_p$ and $\Delta_e^R = -\Delta_p^R$ (see Fig.5). Here, the echo signal will be generated with opposite spectral detuning to the atomic frequency $\omega_{31}$ in comparison with the probe SPP field similar to REQM in a free space scheme [17].

4) $\Omega_{c,2}(-\tau_e) = -\Omega_{c,1}(\tau)$. The concrete calculation of the echo emission shows that we don't need in the $\pi$-shift for realization of efficient echo emission.

5) Finally we note that echo signal emission will evolve reversely in time in comparison with the probe SPP pulse absorption if only the initial conditions of atomic coherences for "P-" and "E-" systems coincide with each others at some fixed moment of time t'. This leads to additional conditions for the excited atomic coherences: a) $\hat{S}_{p13}^j(\tau') = \hat{S}_{e13}^j(\tau_e')$ and b) $\hat{S}_{p12}^j(\tau') = \hat{S}_{e12}^j(\tau_e')$. Here, $\hat{S}_{p13}^j(\tau') = \hat{S}_{e13}^j(\tau_e') = 0$ for large spectral detuning $|\Delta_p| \gg \delta\omega$ and a)-condition leads to the phase matching relation $c(\vec{K}_{cp} - \vec{K}_{cp}) = (n_p \omega_p + n_e \omega_e)\vec{e}_x$ depicted in Fig.4: where $n_p$ and $n_e$ are the refractive indexes for the probe and echo SPP fields. In general case all 1)-5) conditions generalize CRIB procedure of the photon echo QM including REQM.

Below we demonstrate a specific possibility of the nano scale REQM realization for homogeneously broadened atomic transitions 1-3 and 1-2, i.e. $\Delta_{p(e)31}^j = \Delta_{p(e)21}^j = 0$. In accordance with previous analysis, we only need to find appropriate conditions for complete absorption of the probe SPP pulse by resonant atoms on the D/NIMM interface. In this case we get the following absorption coefficient $\tilde{\alpha}_p(\nu)$ by assuming that the Rabi frequency $\Omega_{cp}$ is a constant during the SPP field absorption:

$$\tilde{\alpha}_p(\nu) = \alpha_p(\nu) - 2i\varsigma_p = i\frac{(\chi_p \xi_1^p) C_p(\nu)}{\Delta_p} \ln[\frac{1 - C_p(\nu)}{\exp(-2z_o/\xi_1^p) - C_p(\nu)}], \qquad (17)$$

where $C_p(\nu) = [\Delta_p(\Delta_p^R - \nu) - i\Delta_p \gamma_{21}]/|\Omega_{cp}|^2$. By using (17), we numerically calculated the optical density $\tilde{\alpha}_p(\nu) L_x$ ($L_x = 0.1 \cdot l_x^p$) of the atomic system on the D/NIMM interface as a function of frequency $\nu$ and layer thickness $z_o$ (see Fig.6). The optical density demonstrates a large absorption $\tilde{\alpha}_p(\nu) L_x > 3$ within spectrum range $\Delta_{in} \approx 0.8 \cdot 10^7$ for $z_o > 0.2\lambda_o$ that corresponds to the perfect storage of the SPP field with temporal durations ~0.1 μs for the reasonable atomic parameters in the crystals doped by rare earth ions [27]. By taking into account $\gamma_{21} \approx 10^3 - 10^4$ we get a possibility to store more than 100 light field modes (photon qubits). Further procedure of the QM Protocol contains the use of readout second control SPP field with revised frequency shift ($\Delta_e = -\Delta_p$ in Figs. 4,5) that will lead to the perfect retrieval of the multi mode SPP field in backward "–x" direction to the input probe field.

**Discussion and conclusion**

We found spectral conditions of subwavelength confinement which is accompanied by the low losses of magnetic SPP modes characterized by enhanced electric field component on the D/NIMM interface. The obtained properties of the SPP modes (see Figs. 1, 2) have been also reproduced numerically for larger $\lambda_o$ (around one order of magnitude) that is possible for NIMMs characterized by smaller $\omega_e$ in (4). We note that optimization of 3D spatial design of the D/NIMM interface could open better strategy for realization of 3D confinement and low losses of the SPP fields that is a subject of further studies.

By exploiting the SPP modes, we proposed a nano scale realization of the photon echo QM with direct Raman transition on long-lived atomic levels (nano scale REQM). The storage volume on the D/NIMM interface is determined by the subwavelength transverse confinement $L_y \times z_o$ (where $z_o \sim \xi_1^p < \lambda_o$) with $L_x \approx 1/\tilde{\alpha}_p(\nu)$ along the SPP field propagation. As seen in Fig.6, the perfect nano scale REQM can be realized within spectral range about $\Delta_{in} \approx 0.7 \cdot 10^7$ induced in the atomic system due to the inhomogenous amplitude of the SPP field for layer thickness $z_o > 0.2\lambda_o$ with a minimal optical density $\tilde{\alpha}_p(\nu)L_x > 3$. This is possible due to a considerable enhancement of the interaction between the SPP field and atoms localized near the interface. Here, the atoms feel different additional Stark shifts because of the inhomogeneous spatial profile of the control SPP fields, so the usual CRIB procedure of the frequencies inversion is realized automatically for homogeneously broadened atomic transition that opens a new possibility for such kind of multi mode QM in the nano scale schemes.

**Acknowledgment**

The authors thank the grant of the Russian Foundation for Basic Researches number 08-07-00449, Government contract of RosNauka 02.740.11.01.03 and ESM thanks the grant of Russian Federation President MK 4090.2009.2.

**References:**

[1] T.C. Ralph, *Rep. Prog. Phys*. **69**, 853 (2006).
[2] P. Kok, et al., *Rev. Mod. Phys.* **79**, 135 (2007).
[3] H.-J. Briegel, W. Dur, J. I. Cirac, and P. Zoller, *Phys. Rev.Lett.* **81**, 5932 (1998).
[4] Y. Vlasov, W. M. J. Green and F. Xia, *Nat. Phot.* **2**,2 (2008).
[5] M.P. Nezhad, K. Tetz and Y. Fainman, *Opt. Exp.* **12**, 4072 (2004).
[6] A.A. Govyadinov and V.A.Podolskiy, *Phys. Rev. Lett.* **97**, 223902 (2006).
[7] A.V. Akimov, A. Mukherjee, C.L. Yu, D.E. Chang, A.S. Zibrov, P.R. Hemmer, H. Park and M.D. Lukin, *Nature* **450**, 402 (2007).
[8] M. Fleischhauer, A. Imamoglu, & J.P. Marangos, *Rev. Mod. Phys.* **77,** 633 (2005).
[9] K. Hammerer, A.S. Sorensen and E.S. Polzik, *arXiv:*0807.3358v4 [quant-ph].
[10] W. Tittel, M. Afzelius, T. Chaneli`ere, R.L. Cone, S. Kroll, S.A. Moiseev, and M. Sellars, *Laser & Phot. Rev.* DOI10.1002/ lpor. 200810056 (2009).
[11] A. Lvovsky, B.C. Sanders and W. Tittel, *Nat. Phot.* **3,** 706 (2009).
[12] S. G. Johnson and J. D. Joannopoulos, Photonic Crystals: The Road from Theory to Practice (Kluwer, Dordrecht, 2002).
[13] V.M. Agranovich and D. L. Mills, Surface Polaritons (North-Holland, Amsterdam, 1982).
[14] A. Kamli, S.A. Moiseev and B.C. Sanders, *Phys. Rev.Lett.* **101**, 263601 (2008).
[15] S.A. Maier, Plasmonics: Fundamentals and Applications (Springer, New York, 2007).
[16] S.A. Moiseev, A.A. Kamli and B.C. Sanders, *arXiv:* 0911.1372 [quant-ph].
[17] S.A. Moiseev, W. Tittel, *arXiv:* 0812.1730 v2, [quant-ph].
[18] G. Hetet et al., *Opt. Lett.* **33**, 2323 (2008).
[19] T. van der Sar, E.C. Heeres, G.M. Dmochowski, G.de Lange, L. Robledo, T.H. Oosterkamp, and R. Hanson, *Appl. Phys. Lett.* **94**, 173104 (2009).
[20] S.A. Moiseev, and S. Kröll, *Phys. Rev. Lett.* **87**, 173601 (2001).
[21] S.A. Moiseev, and B.S. Ham, *Phys. Rev. A.* **70**, 063809, (2004).
[22] S.A. Moiseev, V.F.Tarasov and B.S. Ham, *J. Opt. B: Quant. Semicl. Opt.* 5, S497 (2003).
[23] B. Kraus et al., *Phys. Rev. A.* **73**, 020302 (2006).
[24] A.L. Alexander et al., *Phys. Rev. Lett.* **96**, 043602 (2006).
[25] M. Hosseini, et al., *Nature* **461**, 241 (2009).
[26] H.de Riedmatten et al., *Nature* **456**, 773 (2008).
[27] A.V. Turukhin, V S. Sudarshanam, M.S. Shahriar, J.A. Musser, B.S. Ham and P.R. Hemmer, *Phys. Rev. Lett.* **88**, 023602 (2001).

**Figure captions:**

Figure 1. Propagation length of SPP field in units of wavelength $l_x/\lambda_o$ as a function of NIMM permittivity $\varepsilon_2 = \varepsilon_r + i\varepsilon_{im}$ for $\mu_2 = 0$ and dielectric parameters $\varepsilon_1 = 1.31$, $\mu_1 = 1$.

Figure 2. Subwavelength transverse spatial confinement $\xi_1/\lambda_o$ of the SPP field (in units $/\lambda_o$) on the dielectric-NIMM interface as a function of metamaterial permittivity $\varepsilon_2 = \varepsilon_r + i\varepsilon_{im}$ for $\mu_2 = 0$ and for dielectric parameters $\varepsilon_1 = 1.31$, $\mu_1 = 1$. Blue (solid) line - $\varepsilon_{im} = 10^{-3}$, black (dashed) line - $\varepsilon_{im} = 10^{-2}$, red (dotted-dashed) line - $\varepsilon_{im} = 10^{-3/2}$. It is seen that the confinement decreases with losses.

Figure 3. Surface plasmon polariton (SPP) field on the interface of dielectric ($\varepsilon_1 > 0$, $\mu_1 > 0$) and metamaterial ($\varepsilon_2 < 0$, $\mu_2 < 0$), $\upsilon_g$ is a group velocity; $\xi_1$ and $\xi_2$ are spatial confinements of the SPP field near the interface in the dielectric and in the metamaterial; here, we have large confinement $\xi_1 \cong \xi_2 = \xi \ll \lambda_o$, $\lambda_o$ is a wavelength of a free light field in the dielectric; white spots indicate the resonant atoms in the layer with thickness $z_o$.

Figure 4. Spatial scheme of excitation of the probe (echo) and two control SPP pulses with wave vectors $\vec{K}_p$ ($\vec{K}_e$) and $\vec{K}_{cp}$ ($\vec{K}_{ce}$) respectively. Orientation of the wave vector $\vec{K}_{ce}$ is determined by the phase matching condition.

Figure 5. Temporal diagram of the Raman interaction between two pairs of the SPP fields with three-level atomic system. Insert demonstrates the temporal shapes of the SPP pulses and temporal reversibility of the input probe pulse absorption and of the echo signal emission.

Figure 6. Optical density $\tilde{\alpha}_p(\nu)L_x$ ($L_x = 0.1 \cdot l_x^p$) as a function of frequency $\nu$ and of atomic layer thickness in the wavelength unit $z_o/\lambda_o$ for the D/NIMM interface parameters: $\varepsilon_2 = -1.34 + i\,10^{-4}$, $\mu_2 = 0$, $\omega_e = 1.37 \cdot 10^{16}$, $\omega_\mu = \omega_e/1.67$, $\varepsilon_1 = 1.31$, $\mu_1 = 1$; and the atomic parameters: $n_o = 2 \cdot 10^{19}$ cm$^{-3}$, $d_{13} = 10^{-3}$ e $a_0$ (e is electron charge, $a_0$ – Bohr radius), $\Omega_{cp} = 10^7$, $\Delta_p = 10^7$, $\Delta_p^R = 7 \cdot 10^7$, $\gamma_{21} = 10^4$, $\xi_1^p = \lambda_o/40$, $\lambda_o = 285$ nm, $L_z = 0.55\,\lambda_o$. Enough optical density $\tilde{\alpha}_p(\nu)L_x > 3$ occurs for large subwavelength confinement within spectrum width $\Delta_{in} \approx 0.7 \cdot 10^7$.

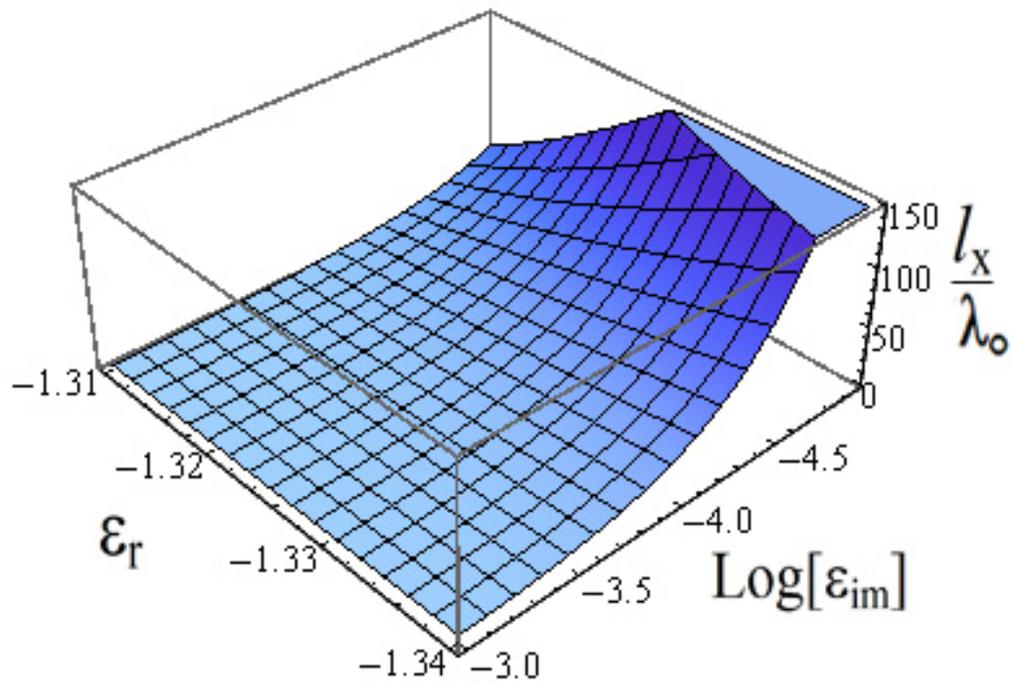

Figure 1.

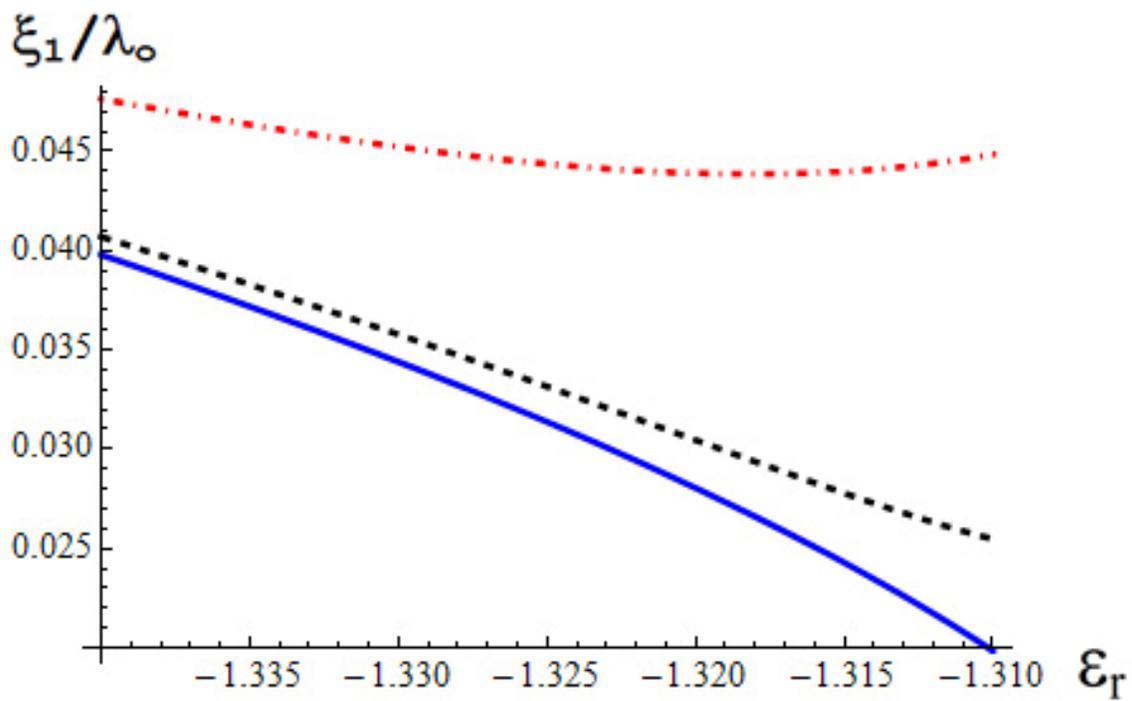

Figure 2.

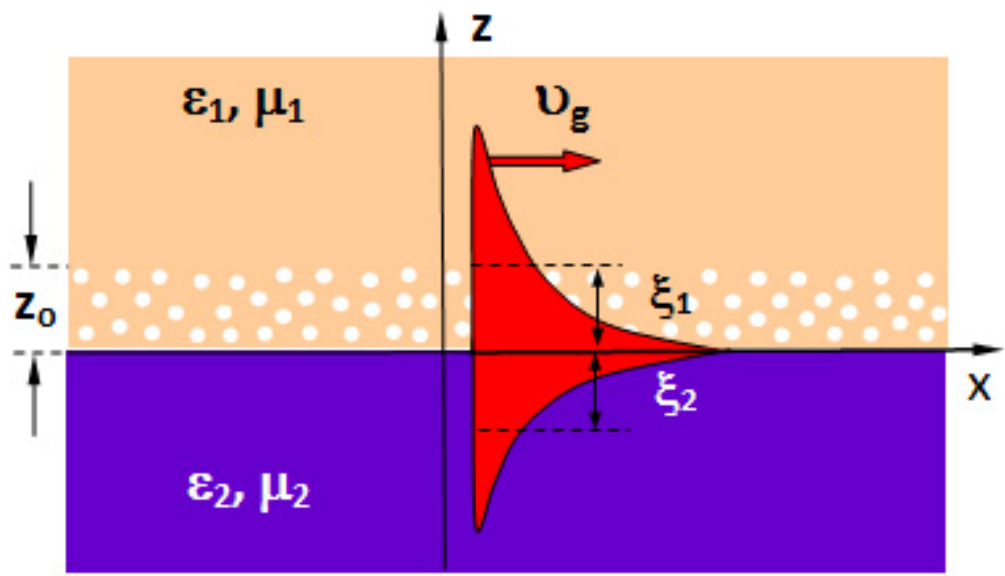

Figure 3.

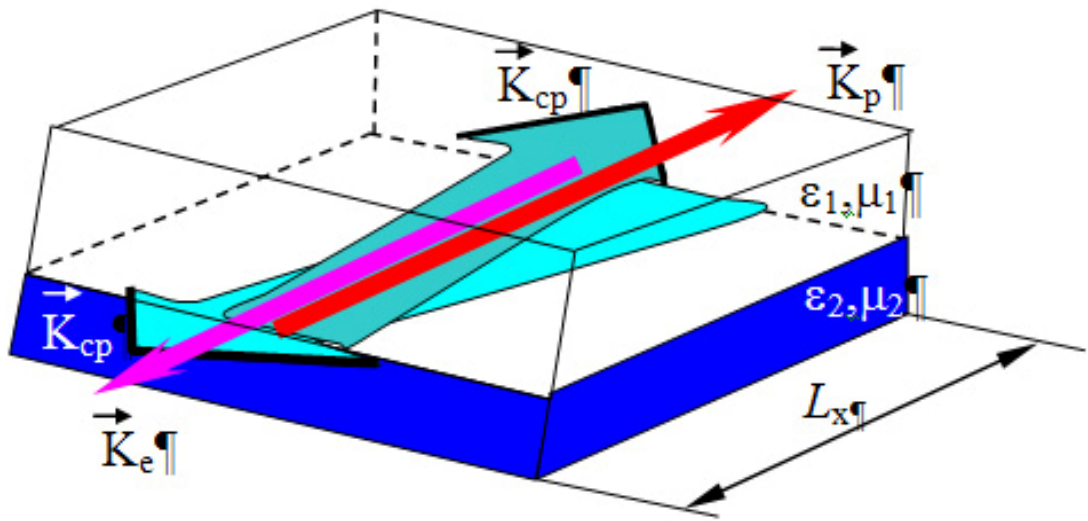

Figure 4.

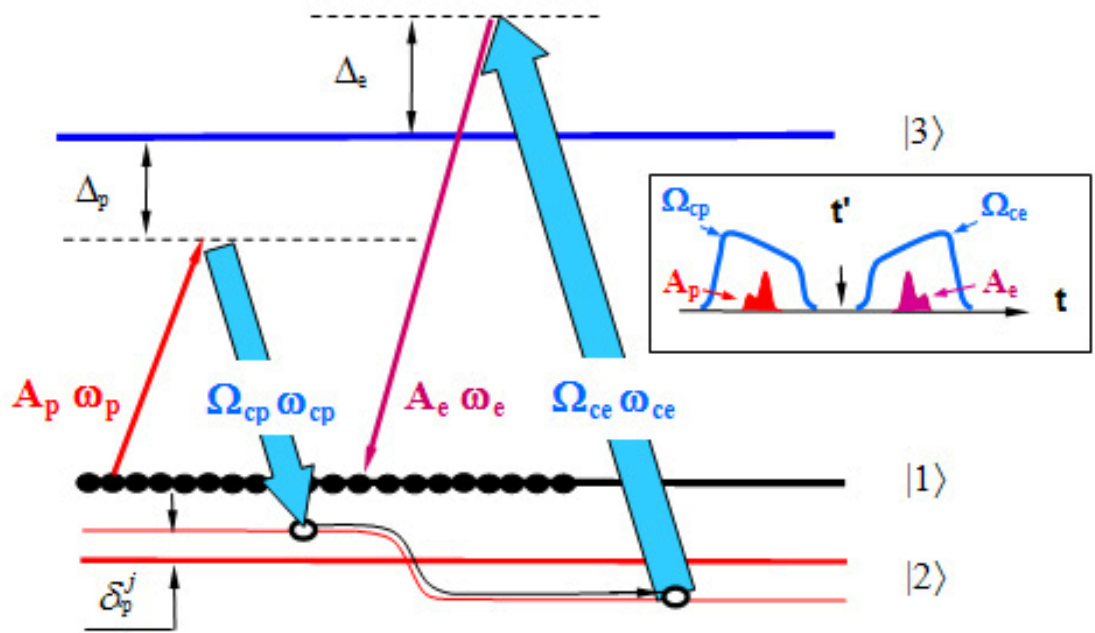

Figure 5.

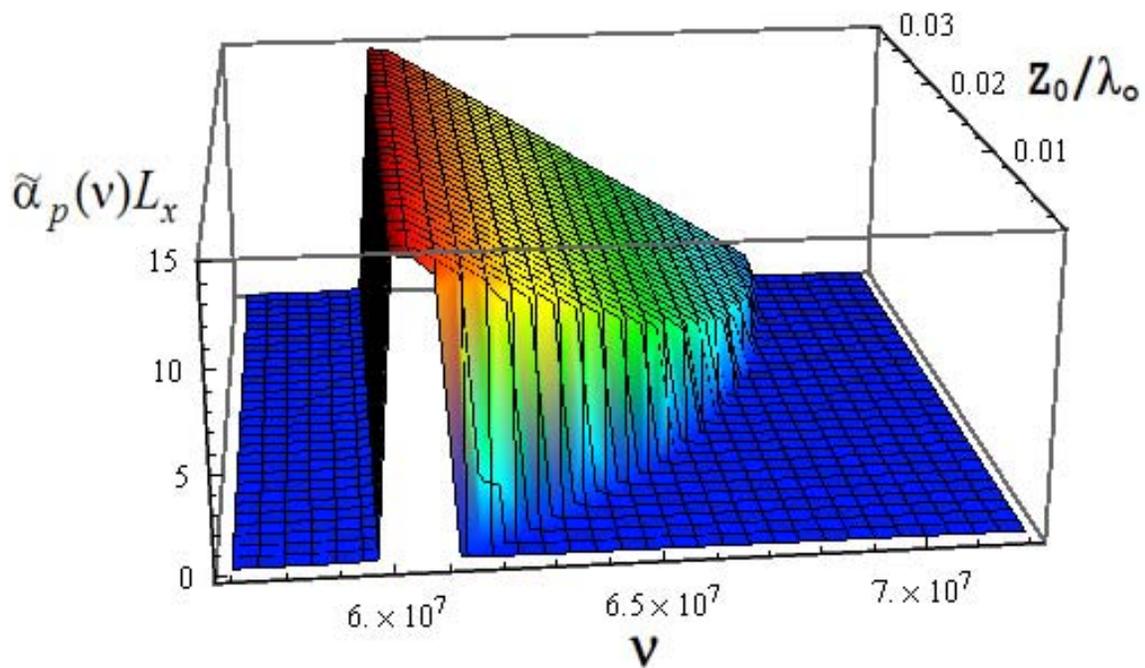

Figure 6.